\documentclass[aps,pra,reprint,groupedaddress,showpacs]{revtex4-1}
\usepackage{amsmath}
\usepackage{amsfonts}
\usepackage{graphicx}
\usepackage{subfigure}
\usepackage{amssymb}

\newcommand{\tr}{{\rm tr}}

\def\lsim{\mathrel{\rlap{\lower4pt\hbox{\hskip1pt$\sim$}}
    \raise1pt\hbox{$<$}}}                
\def\gsim{\mathrel{\rlap{\lower4pt\hbox{\hskip1pt$\sim$}}
    \raise1pt\hbox{$>$}}}                

\begin{document}

\title{Ergodicity from Nonergodicity in Quantum Correlations of Low-dimensional Spin Systems}

\author{R. Prabhu}
\author{Aditi Sen(De)}
\author{Ujjwal Sen}

\affiliation{Harish-Chandra Research Institute, Chhatnag Road, Jhunsi, Allahabad 211 019, India} 


\begin{abstract}
Correlations between the parts of a many-body system, and its time dynamics, lie at the heart of sciences, and they can be classical as well as quantum. 
Quantum correlations are traditionally viewed as constituted out of classical correlations and magnetizations. While that of course remains so,
we show that quantum correlations can have statistical mechanical properties like ergodicity, which is not inherited from the 
corresponding classical correlations and magnetizations, for the transverse anisotropic quantum \(XY\) model in one-, two-, and quasi two-dimension, 
for suitably chosen transverse fields and temperatures. 
The results have the potential for applications in decoherence effects in realizable quantum computers.
\end{abstract}

\maketitle


\section{Introduction}

Quantum entanglement \cite{Horodecki09},  a  striking property of quantum theory, plays a crucial role in many quantum information processes including 
achieving secure quantum cryptography \cite{Gisin02}, faster quantum computers, and better quantum communication \cite{Nielsen00}. 
Quantum entanglement in many-body systems has 
created
considerable interest recently because of its potential 
in understanding quantum many 
body phenomena, which are not amenable to classical perceptions, such as quantum phase transitions \cite{Schadev99}, decoherence \cite{Zurek03}, etc. Also, 
quantum entanglement forms a central element in bringing together many-body physics and quantum information science, and hence in the development of new 
exciting physics \cite{Lewenstein07,Amico08,Bloch08}. Properties of entanglement in many-body systems like cold atomic gases in optical lattices \cite{Jaksch98}, 
 trapped gaseous Bose-Einstein condensates \cite{Duan00},
 and spin models \cite{Osterloh02,Osborne02,Wootters02,Lewenstein07,Amico08} 
have been dealt with increasing interest in recent times.

The importance of studying the time-evolved state cannot be over-emphasized in both many-body physics and quantum information. In particular,
important processes in quantum information, like the one-way quantum computer crucially depends on the time-evolved state and the possibility to control
it in experimental set-ups \cite{NP-Briegel}. 
On the other hand, the changes happening in several physical parameters, including statistical mechanical properties of such parameters,
 during evolution, is an important aspect in many-body physics, to understand, for example, the phenomena of decoherence in the system \cite{Zurek03}.
However, despite its vital importance, the complexity of the time-dynamics in physically interesting models, 
has led to a limited amount of  study of
such phenomena \cite{Lewenstein07,Amico08}.
 
%
The validity of a statistical mechanical description 
of a physical quantity depends on the behavior of that quantity in the time-evolution of  the system. Ergodicity is a  necessary 
condition for the validity of a  statistical mechanical description of a physical quantity. A physical property is said to be 
ergodic if the time average of the quantity matches its ensemble average. 

A two-party quantum state (pure or mixed) has non-zero classical correlations whenever 
it has non-zero quantum correlation with respect to any ``good'' entanglement measure \cite{HHH-limits} (cf. \cite{qcwtcc}). 
However we show that the properties of quantum correlations may not be traced back to those in classical correlations. In particular, 
in the case of 
bipartite quantum states, which in our case are nearest neighbor states of 
quantum spin systems, we show that statistical mechanical 
properties like ergodicity can appear in quantum correlation even when it is absent for all classical correlations and  magnetizations.

To deal with questions about ergodicity, we consider the anisotropic transverse quantum $XY$ model
in one-dimension, ladder, and two-dimension. 
Spectacular advances in cold gas experimental techniques have made these  models experimentally realizable in such systems \cite{Lewenstein07, Bloch08}. 
%
The two-component Bose-Bose and Fermi-Fermi mixture, in the strong coupling limit with suitable tuning of scattering length and additional 
tunneling in the system can be described by the quantum \(XY\) Hamiltonian. 
The dynamics of the system can be simulated by controlling the system parameters and the applied transverse field. 
Therefore the phenomenon discussed in the paper can potentially be verified in the laboratory. 
Moreover, many solid state compounds can be described well by this model. In particular, the
\(\mbox{Li}\mbox{Ho}_x\mbox{Y}_{1-x}\mbox{F}_4\) compound is known to be described by the 
three-dimensional (quantum) transverse  Ising model, which is a specific case of the \(XY\) model \cite{solid-ladder}.

The question of ergodicity in the quantum XY chain was already considered by Mazur in Ref. \cite{Mazur69} for magnetization, and important
further developments were carried out in Refs. 
\cite{Barouch1075,Barouch786} for magnetization as well as classical correlations, and all these physical quantities were shown to be nonergodic. 
Here we show that the corresponding quantum correlation aka entanglement can however be ergodic. Moreover, we carry over the considerations to other low-dimensional spin systems on ladders  and in two-dimensions, and in all cases we show that quantum correlation can be ergodic without the classical correlations and magnetizations being so.

Quantum correlations are known to provide us with applications in quantum information processing tasks, like quantum communication, quantum cryptography, and quantum computation, that cannot be matched by systems with classical correlations only \cite{Horodecki09, Gisin02, Nielsen00}. 
Also, some form of quantum correlation can potentially be a ``universal'' detector of quantum phase transitions \cite{Lewenstein07,Amico08}, 
and numerical algorithms inspired by quantum information 
concepts can potentially provide better understanding of many-body systems \cite{Vidal-algo}. 
And yet, statistical mechanical properties like quantum phase transitions can be identified by using \emph{both} classical \cite{Schadev99} and quantum correlations \cite{Lewenstein07,Amico08}. However, the fact that even statistical mechanical 
properties like ergodicity can emerge in quantum correlation of a quantum state without it being present in the classical correlations (and magnetization) had so far eluded us. The findings can potentially be important in several areas in quantum many-body physics and in 
quantum information. In particular, it may turn out to be crucial for studies on the decoherence times in quantum computers \cite{decoherence-natun}. 

The paper is organized as follows. In Sec. \ref{XYmodel}, we give a brief description of the considered model and the 
time-dependent transverse field. In one-dimension (1D), this model can be exactly diagonalized \cite{Barouch1075,Barouch786, LSM61},  
and we perform the corresponding Jordan-Wigner transformation  to arrive 
at the expressions for magnetization,   classical correlations, and  entanglement  for thermal (canonical) equilibrium states as well as for time-evolved states  for an infinite chain. In this paper, we will use logarithmic negativity as the measure of entanglement, and we 
define this in Sec. \ref{entanglement}. In Sec. \ref{sec:inf}, we consider the question of ergodicity in the infinite chain, and 
we show that 
there are situations where magnetization and all classical correlations are nonergodic, while entanglement remains ergodic. 
Additionally, we show that in a large number of the remaining cases, entanglement is ergodic while magnetization and almost all classical correlations are nonergodic.
Entanglement, therefore, follows the \emph{minority verdict} rather than the majority one: Although most of the classical correlations and magnetization are nonergodic, entanglement is actually ergodic. 
In Sec. \ref{sec:finite}, we show that the statistical behavior of entanglement, classical correlations, and magnetization
 for finite $XY$ chains is  similar to that of the 
infinite 1D $XY$ model, even for relatively moderate system size. Specifically, we consider 
finite chains of size  up to $N=12$, with periodic boundary conditions. The qualitative similarity of the results obtained for finite and infinite chains led us to consider the statistical properties of finite spin systems in two-dimensional (2D) and quasi-2D systems with periodic boundary conditions, where an exact diagonalization procedure is not possible in the infinite limit. We find that the ergodicity of entanglement, despite magnetization and all classical correlations being nonergodic, persists in higher dimensions (Sec. \ref{sec:ladder} and Sec. \ref{sec:2d}).
We summarize our results in the final section (Sec. \ref{sec:conclusion}).


\section{The \(XY\) model in the transverse field}
\label{XYmodel}

\subsection{Description of the model}


We consider an interacting spin Hamiltonian with an external magnetic field given by
\[
H = H_{\rm int} - h(t) H_{\rm mag},
\]
where \(h(t)\) is a time-dependent function. To obtain nontrivial effects on the evolution due to the external magnetic field, 
one must choose the magnetic field and other parameters in such a way that the interaction part and the field part of the Hamiltonian do not commute, i.e., 
$[H_{\rm int},\, H_{\rm mag}]\neq 0$. A simple way to obtain this is to choose the field part as 
\[H_{\rm mag} = \sum_i S_i^z\] and the interaction term as 
\[H_{\rm int} = J \sum_{\langle ij \rangle} \left[(1 + \gamma) S_i^x S_{j}^x + (1 - \gamma) S_i^y S_{j}^y\right],\]
with \(\gamma \ne 0\), 
where $J$ is the coupling constant, $\gamma$ is the anisotropy measure, \(S_i^x\), \(S_i^y\) and \(S_i^z\) are one-half of the Pauli spin matrices 
at the $i^{\rm th}$ site, and $\langle ij\rangle$ denotes interactions between all the nearest-neighbor pairs in a lattice. 
We assume  periodic boundary conditions.
Therefore the time-dependent total Hamiltonian that we study in this paper, reads
\begin{equation}
\label{asolH}
H = J \sum_{\langle ij \rangle} \left[(1 + \gamma) S_i^x S_{j}^x + (1 - \gamma) S_i^y S_{j}^y\right] - 
h(t) \sum_i S_i^z,
\end{equation}
and  is called the anisotropic (quantum) \(XY\) model in a transverse field. 

The transverse field applied is in the form of an initial disturbance that subsequently goes to zero for $t\neq 0$:
\begin{equation}
h(t) = \Big\{\begin{array}{c} a,  \quad t \leq 0 \\ 0, \quad t  >  0  \end{array},
\label{Eq:transfield}
\end{equation}
where \(a \ne 0\). Hence we study the system that has evolved with time after an initial disturbance due to the field.

The state that we consider evolves according to the Hamiltonian \(H\), given in Eq. (\ref{asolH}). But it also depends on the initial state, 
from which it starts evolving. We assume that the initial state of the evolution is 
the (thermal) equilibrium state at the initial time, and at absolute temperature \(T\). Let us denote the canonical equilibrium  state at time \(t\) 
as \(\rho_{eq}^\beta\)(t), so that
\[\rho_{eq}^{\beta}(t) = \frac{\mbox{exp}[- \beta H(t)]}{Z}.\]
Here \(Z\) is the partition function, given by
\[Z = \tr[\mbox{exp}[-\beta H(t)]], \]
and \(\beta = \frac{1}{kT}\), where \(k\) is the Boltzmann constant. 
To consider questions about ergodicity, we will be interested in the behavior of the evolved state. The evolution
is governed by the Hamiltonian \(H(t)\), from a given initial state.
Since we will compare the  long-time average of the properties of the
    evolved state, with those of the equilibrium state, it is natural to suppose that 
the initial state  is the equilibrium state at  \(t=0\), so that there is a possibility that the physical quantity 
after a sufficiently long time, will equilibriate. We denote the evolved state by
\(\rho^\alpha(t)\), where the prefix corresponds to the temperature of the initial equilibrium state 
\(\rho^\alpha_{eq}(0) = \rho^\alpha(0)\). 


\subsection{Single- and two-site reduced density matrices}


The main intention in this paper is to study the statistical behavior of single-site magnetization,
 nearest neighbor entanglement and correlations in various spin models. Therefore we now find out the single-site 
and two-site reduced density matrices of the equilibrium and evolved states. 

A general single-site density matrix is given by
\begin{equation*}
\rho^{\textit 1}=\frac{1}{2}I+2\vec M\cdot\vec S,
\label{eq:}
\end{equation*}
where $I$ denotes the $2\times 2$ unit matrix and magnetization is given as $\vec M=\tr[\rho^{\textit 1}\vec \sigma]$, where \(\vec \sigma = \left( \sigma^x, \sigma^y, \sigma^z \right)\)
are the Pauli matrices.
The single-site density matrix can be obtained by tracing out $N-1$ sites from the $N$-site spin states. 
Let us now consider the single-site density matrix for 
the equilibrium state \(\rho_{eq}^\beta(t)\). Due to symmetry, all the single-site density matrices of the equilibrium states are equal.
We will denote them by \(\rho_{eq}^{\textit{1}}(t)\) (hiding the prefix \(\beta\)).
Now \(\rho_{eq}^{\textit{1}*}(t) = \rho_{eq}^{\textit{1}}(t)\), where the complex conjugation 
is taken in the computational basis, which (for each site) is the eigenbasis of the Pauli matrix \(\sigma^z\).
Therefore $$M_{eq}^y(t) = \tr [S^y \rho_{eq}^{\textit{1}}(t)]=0.$$
Moreover the Hamiltonian \(H(t)\) has the global  phase flip symmetry (\([H,\Pi_i S^z_i]=0\)), from which it follows that 
$$M_{eq}^x(t) = \tr [S^x \rho_{eq}^{\textit{1}}(t)]=0.$$ 
Consequently, the single-site density matrix of the equilibrium state reduces to
$$\rho_{eq}^{\textit{1}}(t) = \frac{1}{2}I + 2 M_{eq}^z(t)S^z.$$
The evolved state does not necessarily have the property of being equal to its complex conjugation,
and consideration of the global phase flip symmetry is complicated 
by the fact that the Hamiltonian is explicitly dependent on time.
%
However, using the Wick's theorem, as in \cite{LSM61,Barouch1075,Barouch786},
the single-site density of the evolved state turns out to be of the form
$$\rho^{\textit 1}(t) = \frac{1}{2}I + 2 M^z(t)S^z.$$
So, the single-site (transverse) magnetization of the equilibrium state is 
\begin{equation}
M_{eq}^z(t) = \tr[S^z \rho_{eq}^{\textit{1}}(t)], \nonumber
\end{equation}
while that for the evolved state is
\begin{equation}
M^z(t) = \tr[S^z \rho^{\textit{1}}(t)]. \nonumber
\end{equation}

The general two-site density matrix for the eqilibrium and the evolved state can therefore be written as
\begin{equation}
\rho^{\textit {12}}=\frac{1}{4}\big[I\otimes I+M^z(\sigma^z\otimes I+I \otimes \sigma^z)+\sum_{i,j=x,y,z}T^{ij}(\sigma^i\otimes\sigma^j)\big],\nonumber
\end{equation}
where 
$T^{ij}=\tr[\rho^{\textit {12}}(\sigma^i\otimes\sigma^j)]$ represents two-site correlation functions. Using Wick's theorem, once again we 
can show that the \(yz\) and \(xz\) correlations will vanish for the evolved state. For the equilibrium state, only the diagonal correlations, \(T^{xx}\), \(T^{yy}\), and 
\(T^{zz}\), remain.

\subsection{Two-site reduced density matrix for the 1D infinite spin model}

The Hamiltonian for the 1D infinite  $XY$ spin model with nearest neighbor interaction
is given by
\begin{equation}
H=\frac{J}{4}\sum_{i}\big[(1+\gamma)\sigma^x_i\sigma^x_{i+1}+(1-\gamma)\sigma^y_i\sigma^y_{i+1}-h(t)\sigma^z_i\big].
\label{eq:1Dinf}
\end{equation}
This model is exactly solvable by successive Jordan-Wigner, Fourier, and Bogoliubov transformations \cite{LSM61}. 
Below we obtain the two-site density matrices for the 
evolved and equilibrium states of the above Hamiltonian $H$.

\subsubsection{Evolved state}

Suppose that the quantum system described by the Hamiltonian \(\textsl{H}\) 
starts off from the initial state which is a (canonical) equilibrium state at temperature \(T\). 
We are interested in the nearest-neighbor (two-site) density matrix of the evolved state at time \(t \rightarrow \infty\), 
that started off from the equilibrium state. 

The correlations and transverse magnetization in this case are given as follows \cite{LSM61,Barouch1075,Barouch786}. 
\begin{eqnarray}
T^{xx}&=&G(-1),\nonumber\\
T^{yy}&=&G(1),\nonumber\\
T^{zz}&=&[M^z]^2-G(1)G(-1)\nonumber\\
T^{xy}&=&T^{yx}=0,
\label{eq:Tijevol}
\end{eqnarray}
where 
$G(R)$ (for $R =\pm 1$) 
is given by
%
%
%
\begin{eqnarray}
G(R) &=&  \frac{1}{\pi}\int^\pi_0d\phi \frac{\tanh(\beta \Lambda(\tilde{a})/2} {\Lambda(\tilde{a})\Lambda^2(0)} (\gamma \sin(\phi R)\sin \phi - \cos^2\phi) \nonumber\\
& &\times (\gamma^2 \sin^2\phi +(\cos\phi-\tilde{a})\cos\phi).
\end{eqnarray}
The transverse magnetization is given by 
\begin{eqnarray}
M^z &=& -\frac{1}{\pi} \int_0^\pi d\phi \frac{\tanh(\beta \Lambda(\tilde{a})/2}{\Lambda(\tilde{a}) \Lambda^2(0)} \nonumber \\
 &\times & \cos\phi [(\cos\phi - \tilde{a})\cos\phi + \gamma^2 \sin^2 \phi]. \nonumber \\
 \label{eq:Mzevol}
\end{eqnarray}
Here 
\begin{equation}
\Lambda(x)= \left\{\gamma^2\sin^2\phi~+~[x-\cos\phi]^2\right\}^{\frac{1}{2}},
\end{equation}
and 
\begin{equation}
 \tilde{a} = \frac{a}{J}, 
\quad \tilde{\beta} = \beta J.
\end{equation}
Note that \(\tilde{a}\)
and \(\tilde{\beta}\) are  
dimensionless variables.



\subsubsection{Thermal state}


%

The correlations and transverse magnetization, for the thermal equilibrium state at a temperature \(T\), are given as follows \cite{LSM61,Barouch1075,Barouch786}. 
The only correlations that remain are the diagonal ones.
\begin{eqnarray}
T^{xx}(a, \beta)&=&G(-1, a, \beta),\nonumber\\
T^{yy}(a, \beta)&=&G(1, a, \beta),\\
T^{zz}(a, \beta)&=& [M^z(a, \beta)]^2-G(1, a , \beta)G(-1, a, \beta),\nonumber
\label{eq:Tijeq}
\end{eqnarray}
where 
$G(R, a , \beta)$ (for $R =\pm 1$) are given by
%
%
%
\begin{eqnarray}
G(R, a , \beta) &=&  \frac{1}{\pi}\int^\pi_0d\phi \frac{\tanh(\tilde{\beta} \Lambda(\tilde{a})/2)} {\Lambda(\tilde{a})} \nonumber\\
&\times &(\gamma \sin(\phi R)\sin \phi - \cos\phi (\cos\phi -\tilde{a})) \nonumber\\
\end{eqnarray}
and 
\begin{eqnarray}
M^z(a, \beta)) &=& -\frac{1}{\pi} \int_0^\pi d\phi \frac{\tanh(\tilde{\beta} \Lambda(\tilde{a})/2) (\cos\phi -a)}{\Lambda(\tilde{a})}. \nonumber \\
 \label{eq:Mzeq}
\end{eqnarray}



\section{Measure of entanglement: Logarithmic negativity}
  \label{entanglement}

  

The bipartite entanglement measure that we will consider throughout the paper is the 
%
   logarithmic negativity (LN) \cite{VidalWerner}. 
It should be stressed, however, that the results do not depend on the choice of the entanglement measure. 
To define logarithmic negativity, let us first introduce negativity.  The negativity \(N(\rho_{AB})\) of a bipartite state \(\rho_{AB}\) 
is defined as the absolute value of the sum of the negative
eigenvalues of \(\rho_{AB}^{T_{A}}\), where \(\rho_{AB}^{T_{A}}\) denotes the partial transpose of \(\rho_{AB}\) with respect to the \(A\)-part \cite{Peres_Horodecki}. 
The logarithmic negativity is defined as
\begin{equation}
  E_{N}(\rho_{AB}) = \log_2 [2 N(\rho_{AB}) + 1].
\label{eq:LN}
\end{equation}
  In our case, the bipartite states are states of two qubits, so that \(\rho_{AB}^{T_{A}}\) has at most 
one negative eigenvalue \cite{Anna-ek}. Moreover, for two-qubit states, a positive LN implies that the state is 
entangled and distillable \cite{Peres_Horodecki, Horodecki_distillable}, while  \(E_{N} =0\) implies that the state is separable \cite{Peres_Horodecki}.


\section{Infinite Quantum $XY$ spin chain in a transverse field}
\label{sec:inf}

In this section, we deal with the question of ergodicity of entanglement and other physical quantities for the 
(one-dimensional) infinite $XY$ spin chain in the time-dependent transverse magnetic field, for which the Hamiltonian is given by Eq. (\ref{asolH}), to be 
considered for 1D here. The Hamiltonian depends on several physical parameters: the coupling \(J\), the anisotropy 
\(\gamma\), and the initial transverse magnetic field \(a\). 
To check for ergodicity of a given physical quantity \(Q\), we begin by looking at its behavior in the time-evolution of the system,
given  by a 
set of system parameters (\(J, \gamma, a\)), and where the initial state of the evolution is the 
canonical equilibrium state of the system at the initial time and 
at a given temperature \(T\). We are interested in the time-averaged value of \(Q\) at large times. Let us call it \(Q^\infty (T, a)\), keeping in mind that it also depends on 
the system parameters \(J\) and \(\gamma\). The physical quantity \(Q\) will be called nonergodic for the system under consideration, if 
its large-time value for the canonical states, \(Q^{\rm can}(T{'}, h(t=\infty))\), 
corresponding to any value of temperature, \(T{'}\), in some relevant range of temperature around \(T\), does not match \(Q^\infty(T,a)\). In other words, 
\begin{equation}
\label{raat-duto-egaro}
Q^\infty(T,a) \ne Q^{\rm can}(T{'},h(t=\infty)) \quad \forall T{'},
\end{equation}
where \(T{'}\) is chosen in a physically relevant range,
will imply nonergodicity of \(Q\).
Otherwise, it will be termed ergodic.
We term a physical quantity as strongly nonergodic if the relation (\ref{raat-duto-egaro}) holds for all temperatures \(T{'}\).

We begin by looking at the (transverse) magnetization in this model. The results and discussions, along with the figures presented, are 
for \(\gamma = 1/2\). However, all the results are independent of the value of \(\gamma\).  
Choosing the temperature of the initial canonical state at an exemplary value, given by 
\(\tilde{\beta} =20\),
we find the long-time average of magnetization in the 
time-evolved state of the system for different values of the initial magnetic field \(a\). 
As exemplary cases, 
for \(a/J = 0.2\), \(0.6\), and \(2\), 
the long-time averages of the transverse magnetizations
in the time-evolved states,
where the dynamics is assumed to start off from the canonical equilibrium state with \(\tilde{\beta} =20\),
are \(0.005\), \(0.079\), and \(0.643\).
Actually,  these long-time averages of magnetization are all nonzero, as long 
as the initial magnetic field is nonzero. The canonical magnetization (i.e., the magnetization in  the canonical equilibrium state) is however zero 
for all nonzero times, 
and at all temperatures, 
implying that the magnetization is
 nonergodic in this model \cite{Mazur69, Barouch1075,Barouch786}. 
The difference between the magnetizations of the evolved and the equilibrium states directly
 depends on the applied field strength.

The situation for the
nearest neighbor correlations,  \(T^{xx}\), 
\(T^{yy}\), and \(T^{zz}\), is a bit more involved. 
For low values of the initial magnetic field, the correlations remain ergodic. 
As we crank up the initial transverse field, the \(yy\) and \(zz\) fields behave similarly, and become strongly nonergodic, just like 
the transverse magnetization: There is no temperature for which the corresponding equilibrium values match the long-time averages. 
The situation for the \(yy\) correlation is shown in Fig. 1 (middle).
The \(xx\) correlation is ergodic for low magnetic fields. However, for high magnetic fields, \(T^{xx}\) attains 
ergodicity only if we allow temperatures much higher as compared to the initial temperature. See Fig. 1 (top). 
For example, for \(a/J=2\), the intersection (of the equilibrium curve and the time average horizontal line) is at \(\tilde{\beta} <0.5 \), which is two orders 
of magnitude below the inital \(\tilde{\beta} =20\). This is not physically meaningful. Therefore, if we suppose that we allow the 
temperature of the equilibrium state to be within about one order of magnitude around the initial temperature, the equilibrium and time-evolved curves for 
\(T^{xx}\) will 
not intersect, and  we obtain nonergodicity of the \(xx\) correlation in this model. 
%

Surprisingly, the situation is the opposite for quantum correlations \cite{amader-purono, abar-purono}. 
The long-time average of the nearest neighbor entanglement  in the time-evolved state (with the 
initial state of evolution being the canonical state with the same temperature as in the case of correlations and magnetization) is decreasing 
with increasing initial magnetic field. And for example, all the three horizontal lines,
in Fig. 1 (bottom),  
for time-averaged entanglement corresponding to \(a/J= 0.2\),  \(0.6\), and \(2\), 
 intersect the curve corresponding to the canonical entanglement at different temperatures (within one order of magnitude of the initial temperature), 
implying that entanglement is actually ergodic in these models. 

Ergodicity can therefore appear in a physical quantity (entanglement, here), without it being present in the constituent physical quantities (the correlations and 
magnetization, here). 

\begin{figure}[h]%
\resizebox{0.6\columnwidth}{!}{
\includegraphics{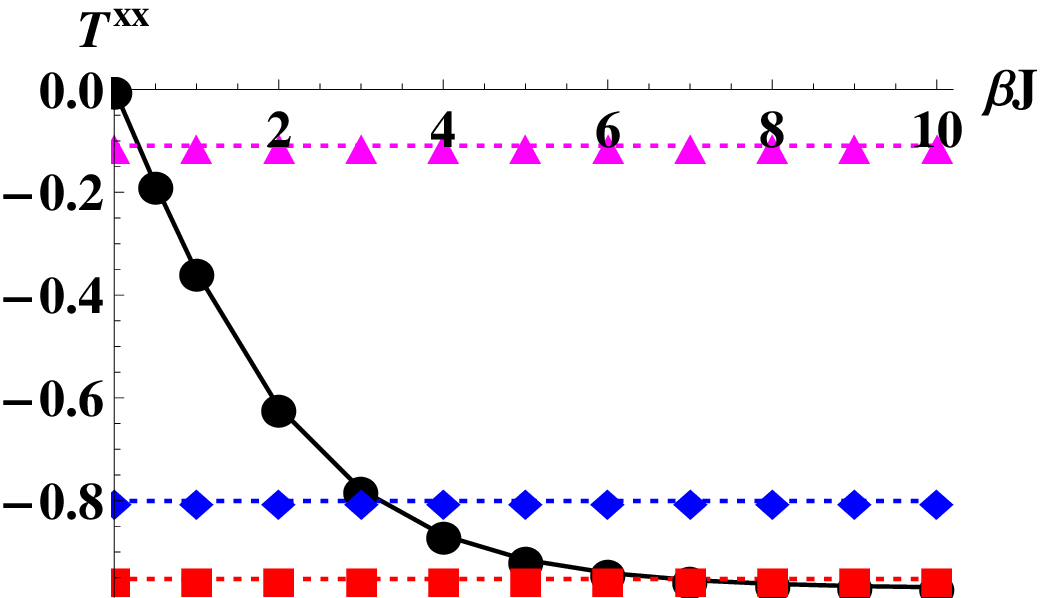}%
}
\resizebox{0.6\columnwidth}{!}{
\includegraphics{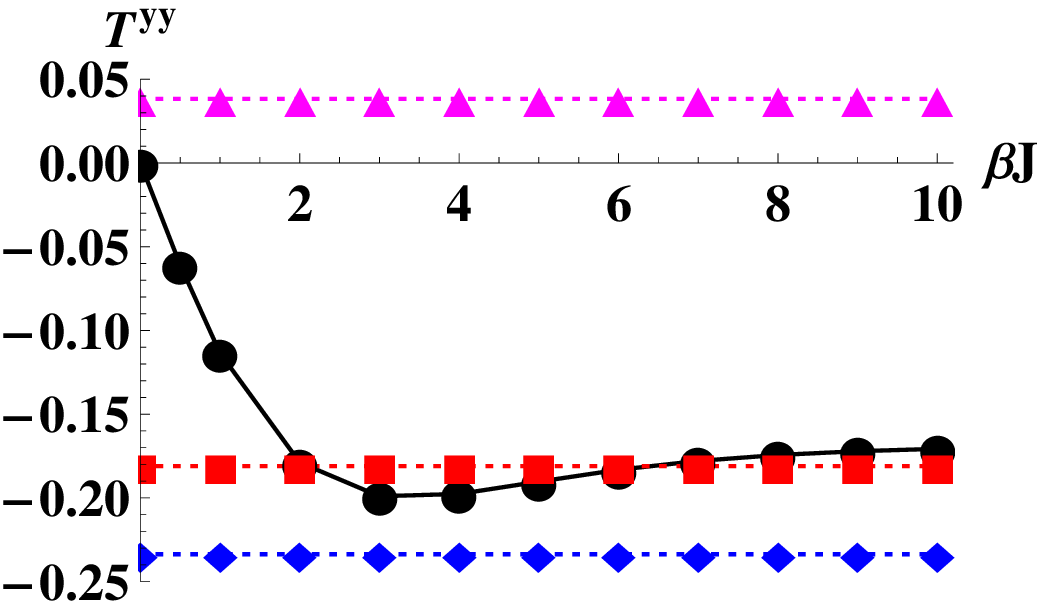}%
}
\resizebox{0.6\columnwidth}{!}{
\includegraphics{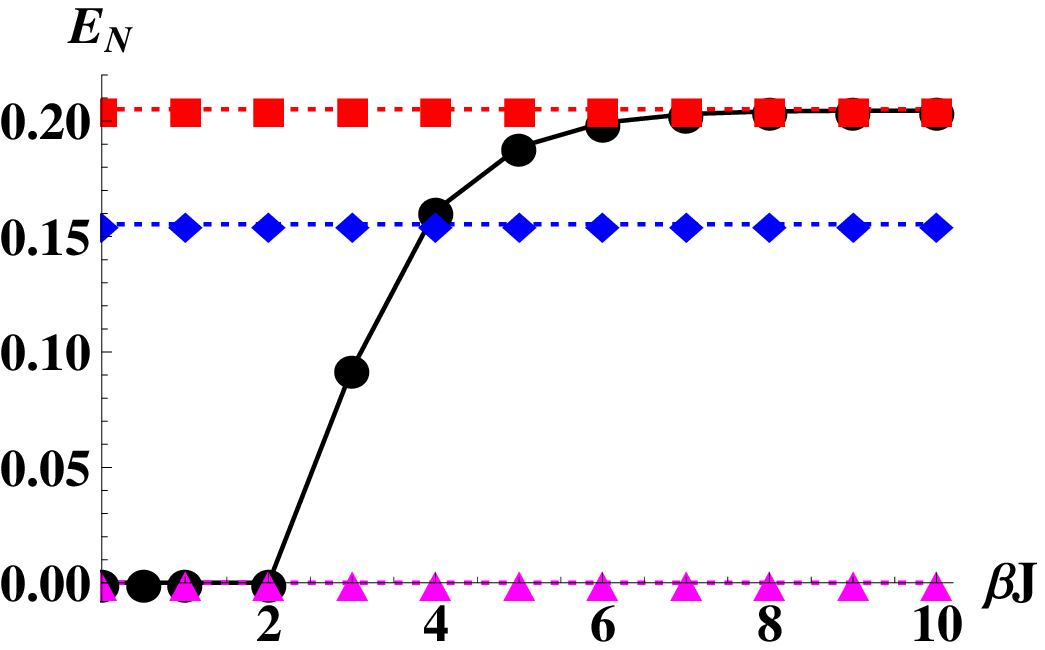}%
}
\caption{(Color online) Correlations (dimensionless) and entanglement (in ebits) 
for the infinite 1D XY model. The horizontal axis represents the dimensionless parameter \(\tilde{\beta}\). 
All the classical correlations are ergodic at low fields. 
The correlations \(T^{yy}\) (middle) and \(T^{zz}\) become nonergodic at high initial fields. 
The \(xx\) correlation (top) is ergodic at low fields, while at high fields, it is nonergodic if we assume that the equilibrium temperature 
can only be an order of magnitude different from the initial temperature.  
The opposite is true for entanglement. 
The curves with black circles represents the equilibrium values as functions of 
\(\beta J\). 
For easy comparison, we have plotted the horizontal lines 
to show the values of the corresponding physical parameter for different values of the initial magnetic field: precisely, for 
$a/J=0.2$ (red squares), 
$a/J=0.6$ (blue diamonds),
and $a/J=2$ (pink triangles). \(a/J\) is dimensionless.
}%
\label{fig:combi1dNinf}%
\end{figure}

\section{Finite $XY$ spin chain in a transverse field}
\label{sec:finite}

Having found the status of the ergodicity of magnetization, and classical and quantum correlations for the transverse \(XY\) chain, we would like to look at its 
character for higher dimensional systems. In these cases, however, there are no analytical diagonalizations possible, and we have to resort to numerical or approximate methods. 
We will be using exact diagonalizations for finite periodic higher dimensional systems, to find the status of ergodicity of magnetization, and classical and 
quantum correlations 
in such systems. But before that, in this section, we consider finite periodic chains, and try to check whether the inferences on ergodicity remains the same
as those from infinite chains.

The Hamiltonian for a (one-dimensional) spin chain of 12 spins, with periodic boundary conditions,  is given by
\begin{equation}
\label{12H}
H = J \sum_{i=1}^{12} \left[(1 + \gamma) S_i^x S_{i+1}^x + (1 - \gamma) S_i^y S_{i+1}^y\right] -  
h(t) \sum_{i=1}^{12} S_i^z,
\end{equation}
Here, we have assumed  periodic boundary conditions, whereby $\vec S_{12+1}=\vec S_1$.

As in the case of the infinite chain, we start the evolution with the canonical equilibrium state at a temperature given by \(\tilde{\beta} =20\), 
and compare the long-time average of a 
certain physical quantity with the values of the same quanitity in the canonical equilibrium state at different temperatures. 

As before, the picture is simple  for magnetization. It is always zero for the equilibrium state at any nonzero time, while it has a nonzero 
time-averaged value in the time-evolved state for any nonzero initial transverse field. Hence, magnetization is strongly nonergodic 
for all nonzero \(a\). The behavior of all the classical correlations in the 
finite chain is also similar to that in the infinite chain.
Of the classical correlations, let us, for definiteness, consider the \(yy\) correlation, \(T^{yy}\). As seen in Fig. 2 (left), it remains ergodic for 
low values of the applied field, while becoming strongly nonergodic for higher values. However, Fig. 2 (right) shows that with increasing applied magnetic field, 
the entanglement becomes ergodic. 

It is clear therefore that already at the level of $N=12$ spins of the one-dimensional $XY$ model, the statistical mechanical 
behavior of nearest neighbor entanglement and classical correlations, and single-site 
magnetization, mimics those of the infinite chain. 
%
This convinces us to study such statistical behavior 
in the case of similar systems in higher dimensions, where choosing a finite number of particles is a necessity.
This will be taken up in the two succeeding sections.


\begin{figure}[h]%
\includegraphics[width=1.75 in]{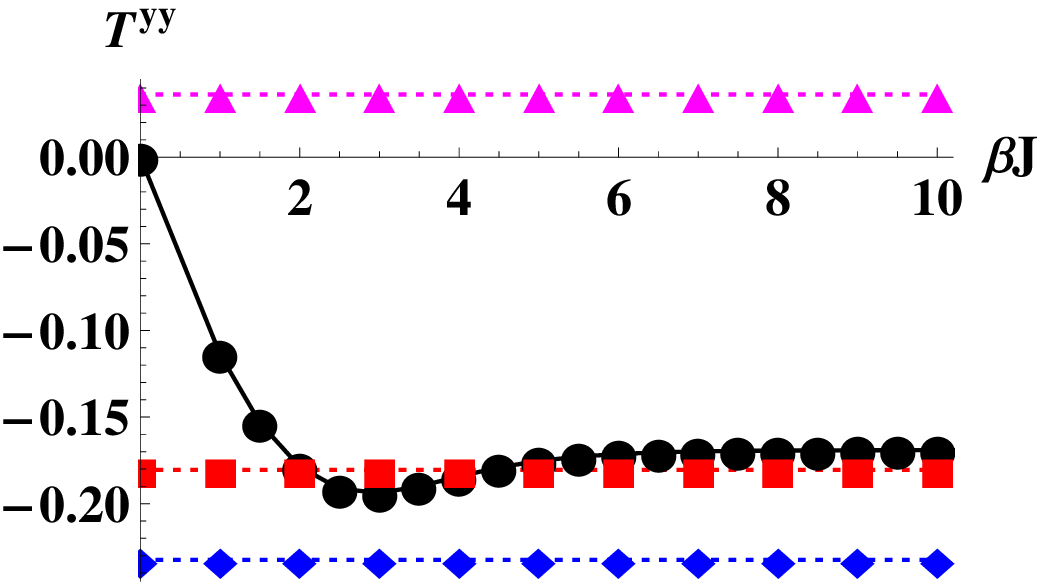}%
\includegraphics[width=1.75 in]{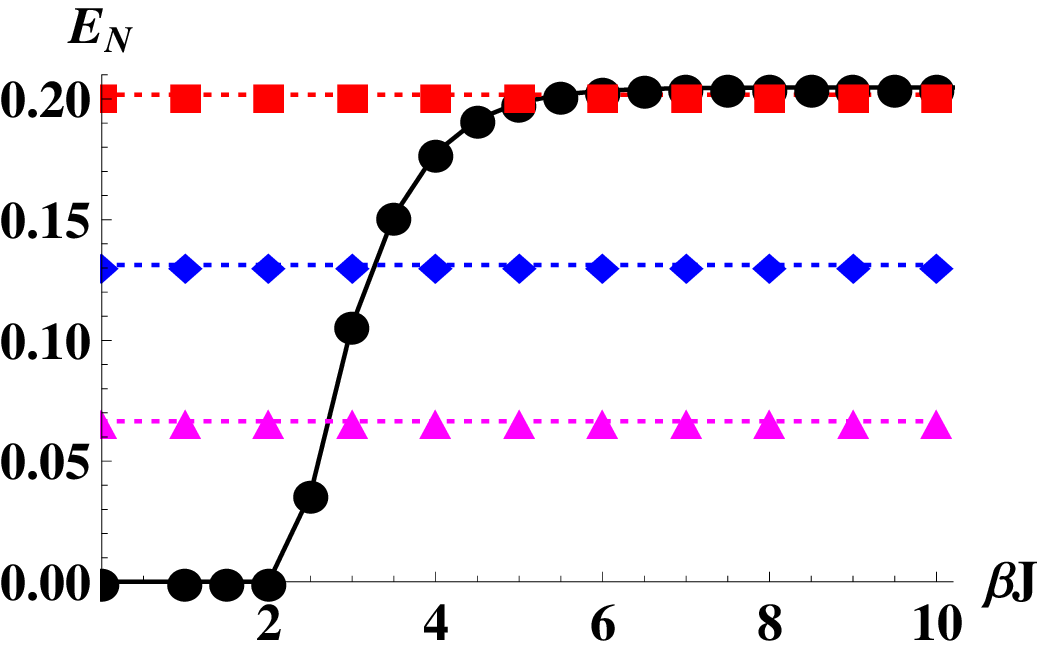}%
\caption{(Color online) Correlation (dimensionless) and entanglement (in ebits) for a chain of 12 spins, with periodic boundary conditions.
The depictions are as before. 
Clearly, the correlation \(T^{yy}\) is ergodic for low magnetic fields, while becoming nonergodic for higher values of the field.
The opposite is valid for entanglement.
}%
\label{fig:combi1dN12}%
\end{figure}

\section{Quasi-2D: $XY$ model with a transverse field on a ladder}
\label{sec:ladder}

A ladder of spins forms an interesting intermediate dimension between one and two dimensions, containing several intriguing features of both the worlds. Moreover, there are materials that have been discovered which can be described by a ladder Hamiltonian. Currently available techniques in 
cold gas systems also offer the possibility of realizing such models \cite{ladder-reference}.


\begin{figure}[h]
\includegraphics[width=\columnwidth]{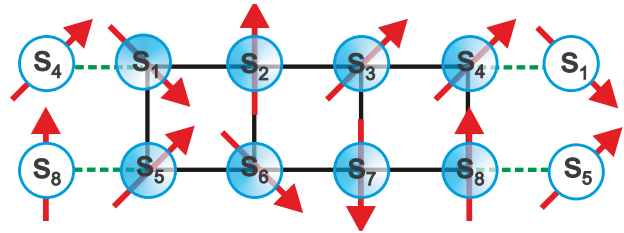}%
\caption{(Color online) A ladder of 8 spins with quantum $XY$ rails and rungs. Periodic boundary condition is assumed along the rails.}%
\label{fig:ladder8}%
\end{figure}

\begin{figure}%
\includegraphics[width=1.75 in]{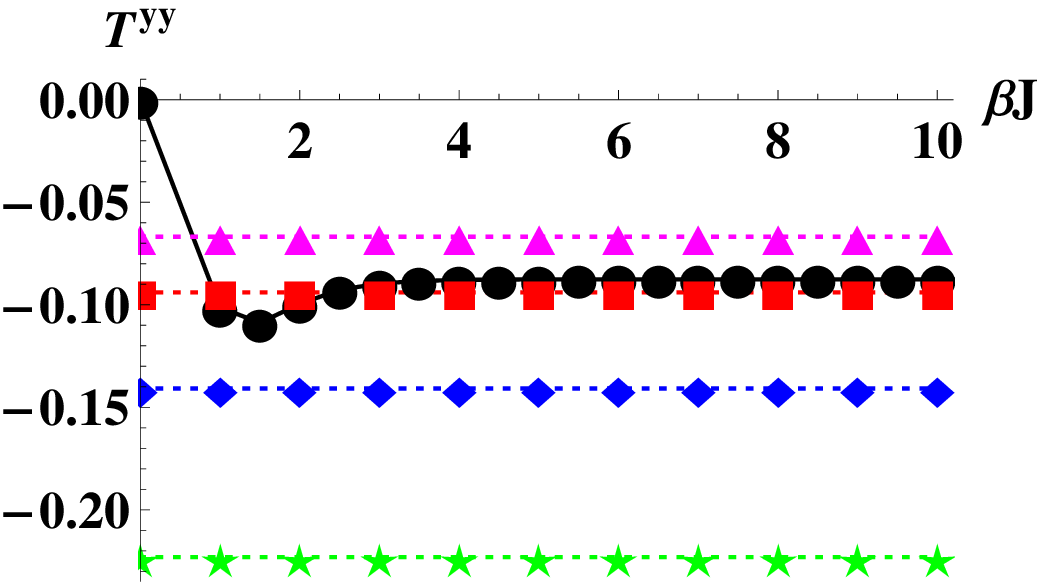}\includegraphics[width=1.75 in]{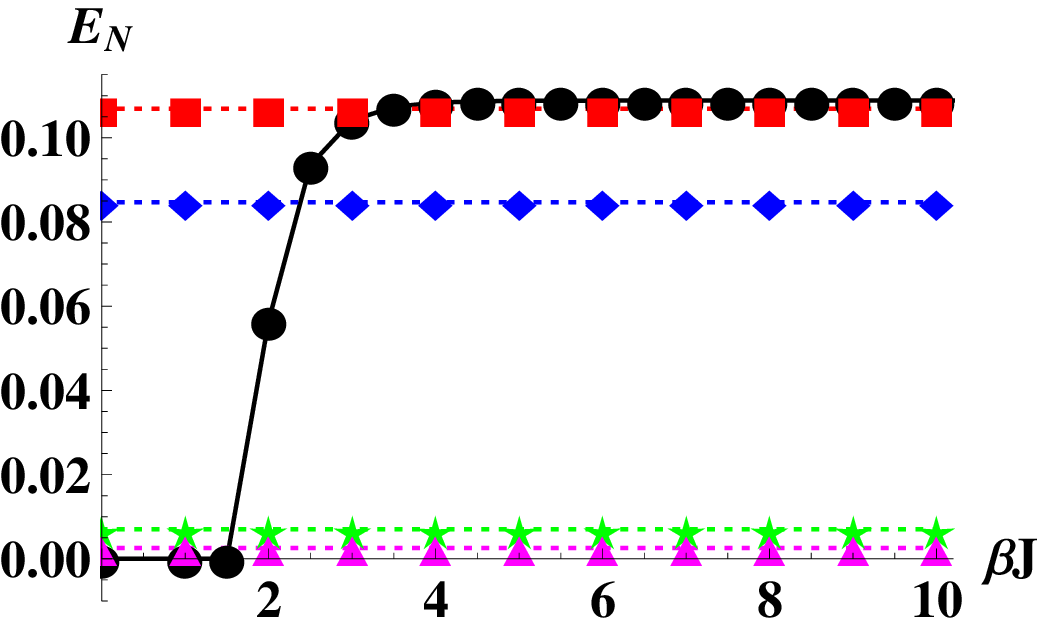}%
%
\caption{(Color online) Correlation (dimensionless) and entanglement (in ebits) on a ladder with periodic boundary conditions. 
Green stars represent \(a/J=1.2\). The other depictions are the same as before. At high magnetic fields, correlations become nonergodic, while 
entanglement is ergodic.
%
}%
\label{fig:combiLadN8}%
\end{figure}

Eight spins arranged in the form of a ladder is shown in Fig. \ref{fig:ladder8}, where we have assumed periodic boundary conditions along the rails, so that 
$\vec{S}_{4}\) and \(\vec{S}_{1}$,  and $\vec{S}_{8}\) and \(\vec{S}_{5}$ are connected by quantum \(XY\) interactions just like all other 
nearest neighbor spins. 
Let us now study 
the 
statistical mechanical behavior of nearest neighbor correlations and entanglement, and single-site magnetization, for such a spin model. 
In Fig. \ref{fig:combiLadN8}, we plot the nearest neighbor 
entanglement and the classical correlation $T^{yy}$, for this system, as a function of temperature for various applied transverse field strengths, 
and where again we assume that the evolution starts off from a temperature given by \(\tilde{\beta} = 20\). 

{\em Magnetization:} The transverse magnetization is clearly nonergodic, as the equilibrium value is zero for all temperatures, while the long-time averaged values
are nonzero for any nonzero applied initial field.

{\em Correlations:} 
For the ladder of 8 spins, 
let us begin by considering the behavior of the $T^{yy}$. See Fig. \ref{fig:combiLadN8} (left). 
The long-time canonical 
equilibrium  $T^{yy}$, is vanishing at zero \(\beta\),
which then decreases to a value of $-0.109$, and converges to \(-0.088\) for low temperatures. 
Now the long-time average value of the evolved state $T^{yy}$, for 
an applied transverse field $a/J= 0.2$ is  $-0.094$,  which is represented by an horizontal line (of red squares) in Fig. \ref{fig:combiLadN8} (left).
It clearly indicates that there exists a temperature for which \(T^{yy}\) of the canonical equilibrium state coincides with the long-time average of \(T^{yy}\) 
of the evolved state, predicting ergodicity of \(T^{yy}\).
%
However, for 
higher values of the applied field strength, the situation is the opposite, and for example, for $a/J=0.6$ and $a/J=1.2$, the respective horizontal lines 
%
do not intersect the equilibrium curve, indicating strong nonergodicity of the $T^{yy}$ correlation for these models. 
For an even higher field of \(a/J=2\), the intersection is at a temperature that is more than an order of magnitude different from the 
initial temperature, and we predict nonergodicity again.
The correlation $T^{zz}$ is ergodic at low fields, and strongly nonergodic for high fields. 
The correlation \(T^{xx}\) is again ergodic at low fields, but at high fields, it is nonergodic.

{\em Entanglement:} The behavior of entanglement is exactly the opposite to that of classical correlations. The long-time average 
of entanglement decreases with increasing magnetic field. 
As seen in Fig. \ref{fig:combiLadN8} (right), for the exemplary values, \(a/J = 0.2, 0.6, 1.2\) and \(2\), of the initial magnetic field, 
the behavior of entanglement
is clearly ergodic.

Thus, it is clear 
that 
for the spin ladder (quasi-2D) model with quantum 
$XY$ rails and rungs, the statistical behavior of nearest neighbor entanglement and classical correlations, and single-site magnetization,
is qualitatively the same  as that of the spin chain: Quantum correlations can be ergodic without classical correlations and magnetization 
being so.

\section{Transverse \(XY\) model in 2D}
\label{sec:2d}

The quantum \(XY\) model in a transverse field in two dimensions is an important system in several applications. In particular, in the special case of unit anisotropy,
the dynamics of this model is utilized in the measurement-based model of quantum computation \cite{NP-Briegel}. 


\begin{figure}[h]
\includegraphics[width=\columnwidth]{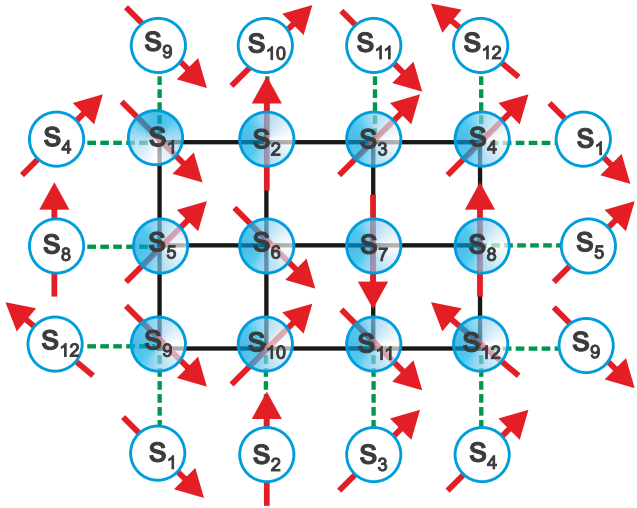}%
\caption{(Color online) A two-dimensional array of 12 spins with $XY$ interactions. Periodic boundary conditions is assumed, so that the system forms a torus.}%
\label{fig:2darray}%
\end{figure}

\begin{figure}[h]%
\includegraphics[width=1.75 in]{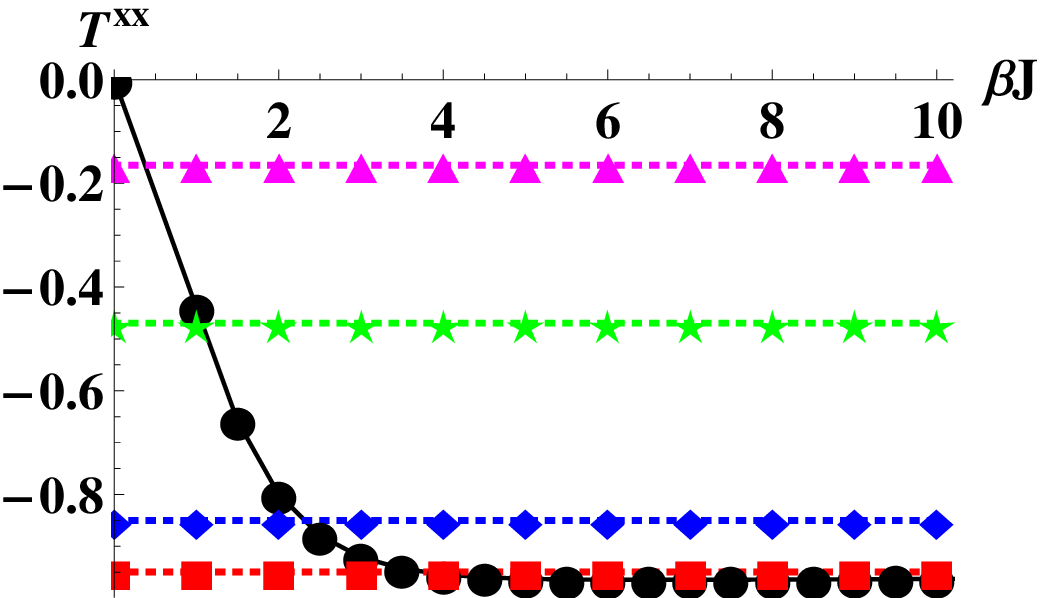}%
\includegraphics[width=1.75 in]{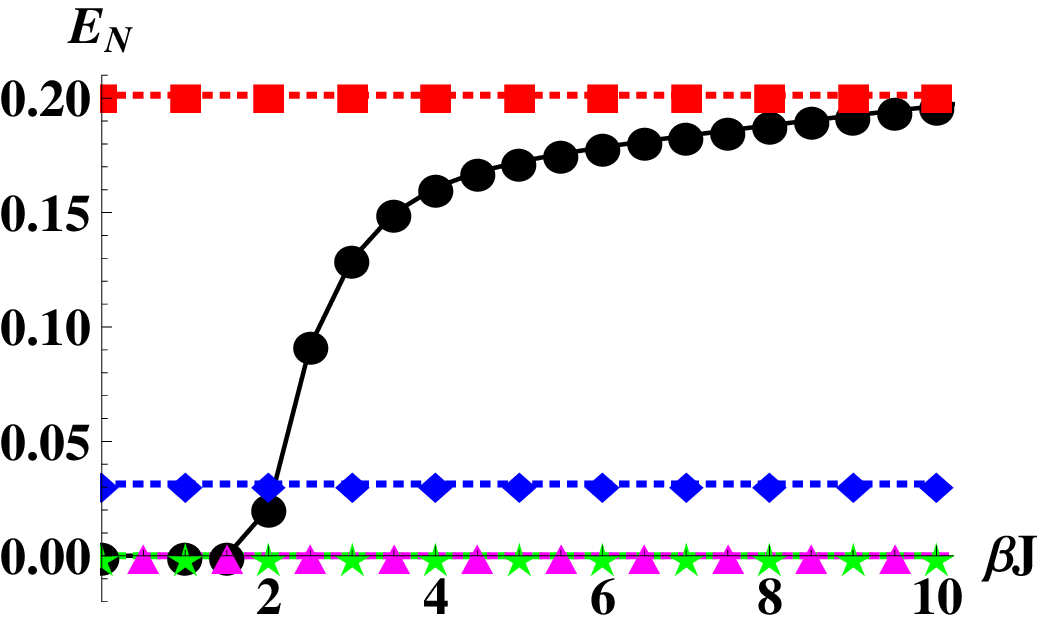}%
%
\caption{(Color online) Correlation (dimensionless) and entanglement (in ebits) on a 2D square lattice. 
The depictions are the same as before. And as before, high magnetic fields render the correlations as nonergodic, while entanglement is ergodic there.
}%
\label{fig:combi2dN12}%
\end{figure}

We consider a  two-dimensional array of 12 spin-1/2 particles arranged on a square lattice, and 
interacting via nearest neighbor interactions. A schematic representation is given in Fig. \ref{fig:2darray}, where periodic boundary condition 
is assumed, so that the system forms a torus. 
We now study the ergodicity properties of this torus for 
nearest neighbor entanglement and classical correlations, and single-site magnetization. In Fig. \ref{fig:combi2dN12}, we plot the nearest neighbor 
entanglement and the $T^{xx}$ correlation for the torus of 12 spins, as a function of temperature for various applied transverse field 
strengths, where again the initial state of the evolution is the canonical equilibrium state at a temperature given by \(\tilde{\beta}=20\). 

{\em Magnetization:} The situation for the transverse magnetization remains the same as in the other dimensions: 
While the canonical equilibrium state has zero magnetization for all temperatures,
the time-evolved state, when averaged over a large time (away from the initial time), produces nonzero magnetization for all nonzero initial fields.

{\em Correlations:} 
The behavior of the $T^{xx}$ correlation is given in the 
Fig. \ref{fig:combi2dN12} (left). The equilibrium state $T^{xx}$ (curve with black circles) is of course vanishing at \(\beta =0\),
which then decreases to $-0.963$ at low temperatures. 
The long-time average of $T^{xx}$ in the time-evolved state (with the initial state of the evolution being the canonical equilibrium state with \(\tilde{\beta} = 20\))
 has different values for different initial applied fields. They are represented as horizontal lines in Fig. \ref{fig:combi2dN12} (left), 
for 
$a/J= 0.2,\, 0.6$, $1.2$, and \(2\).
The qualitative feature is again the same as in the other dimensions: Although
$T^{xx}$ is ergodic for low applied magnetic fields, it ceases to be so for high magnetic fields, although the nonergodicity
obtained is not ``strong''.
The  behavior of the $T^{yy}$ correlation is  similar to that of \(T^{xx}\).
The $T^{zz}$ correlation is ergodic for low field strengths, while becoming strongly nonergodic for high fields.

{\em Entanglement:} We observe that entanglement is ergodic for all values of the applied initial magnetic field. 
In Fig. \ref{fig:combi2dN12} (right), the long-time average of the nearest neighbor entanglement is depicted as horizontal lines for 
\(a/J = 0.2, 0.6, 1.2\) and, \(2\), and all of them intersect the corresponding equilibrium curve.
%

Note that we find that more classical correlations are violating strong non-ergodicity as we increase the dimension of the system.

\section{conclusion and discussion}
\label{sec:conclusion}

We considered the transverse quantum \(XY\) model with nearest neighbor interactions in low dimensions. 
We found that all the two-body classical correlations along with the magnetization in the system, 
with a suitably chosen magnetic field, cannot 
equilibriate,
and hence remain nonergodic, while 
the bipartite entanglement in the system with the same magnetic field can be ergodic. The thesis is true for the infinite chain,
 which is exactly diagonalizable. 
The behavior low dimensional finite systems -- ladder and two-dimensional models with periodic boundary conditions -- mimics the one-dimensional infinite one.

We are therefore led to the conclusion that although quantum correlations have traditionally been understood to be constituted out of classical correlations
and magnetizations, 
it does not necessarily imply that statistical mechanical properties of quantum correlations
will be inherited from those of classical correlations and magnetizations. 

Apart from the above implication of fundamental nature, the result can potentially 
have implications in the decoherence problems in  
quantum information processing tasks \cite{decoherence-natun}. 

Lastly, entanglement has been known to be useful for information processing tasks, and in these tasks, it clearly outperforms all systems which only possess 
only classical correlations \cite{Horodecki09, Gisin02, Nielsen00}. Recent investigations have revealed the importance of entanglement-like concepts 
in dealing with statistical mechanical concepts like quantum phase transitions \cite{Lewenstein07, Amico08}, 
which however, can also be detected by using classical correlations \cite{Schadev99}. The results obtained in this paper point out that even 
statistical mechanical phenomena, like ergodicity, can appear in quantum entanglement independently of classical correlations.





\begin{acknowledgments}

We acknowledge discussions with Micha{\l} Horodecki, and computations performed at the cluster computing facility in HRI (http://cluster.hri.res.in/).

\end{acknowledgments}


\begin{thebibliography}{99}
\bibitem{amader-purono} It was shown previously in Ref. \cite{abar-purono} that entanglement can be nonergodic in this model in 1D, if we restrict attention to the 
energy surface.
\end{thebibliography}
\end{document}